\documentclass[hyper]{prop2015}

\usepackage[english]{babel}
\usepackage{color}

\category{Proceedings}
\subtitle{\href{http://www.maths.dur.ac.uk/lms/109/index.html}{LMS/EPSRC Durham Symposium on Higher Structures in M-Theory}}
\title{Higher Forms and Membranes in 4D Supergravities}
\author[I. Bandos]{I. Bandos\inst{a}}
\author[F. Farakos]{F. Farakos\inst{b}}
\author[S. Lanza]{S. Lanza\inst{c}}
\author[L. Martucci]{L. Martucci\inst{c}}
\author[D. Sorokin]{D. Sorokin\inst{c,}\footnote{Corresponding author e-mail:~\href{mailto:sorokin@pd.infn.it}{\textsf{sorokin@pd.infn.it}}}}
\address[1]{Department of
	Theoretical Physics, University of the Basque Country UPV/EHU,
	P.O. Box 644, 48080 Bilbao, Spain and IKERBASQUE, Basque Foundation for Science, 48011, Bilbao, Spain}
\address[2]{KU Leuven, Institute for Theoretical Physics,  Celestijnenlaan 200D, B-3001 Leuven, Belgium}
\address[3]{Dipartimento di Fisica e Astronomia `Galileo Galilei',  Universit\`a degli Studi di Padova and I.N.F.N. Sezione di Padova, Via F. Marzolo 8, 35131 Padova, Italy}
\shortauthors{I. Bandos, F. Farakos, S. Lanza, L. Martucci, D. Sorokin}

\begin{acknowledgements}
Work of I.B. was supported in part by the
Spanish MINECO/FEDER (ERDF) EU  grant FPA 2015-66793-P, by the Basque Government Grant IT-979-16, and the Basque Country University program UFI 11/55. 
F.F. is supported from the KU Leuven C1 grant ZKD1118 C16/16/005. 
S.L. is grateful for hospitality to Istituto de F\'isica Te\'orica UAM-CSIC, Madrid, where this work was completed.
Work of D.S. was supported in part by the Russian Science Foundation grant 14-42-00047 in association with Lebedev Physical Institute  and by the Australian Research Council project No. DP160103633.
\end{acknowledgements}

\begin{abstract}
  We review the dynamical generation of coupling constants in 4D supergravity by means of gauge three-form fields. 
The latter are introduced as components of particular chiral supermultiplets 
and can be coupled to membranes preserving local supersymmetry. Such a set-up naturally arises from type-II string compactifications on Calabi--Yau manifolds with fluxes.
We present generic 4D $\mathcal N=1$ supergravity models with three-form multiplets and study domain wall solutions supported by membranes,
which interpolate between vacua with different values of the cosmological constant.  
\end{abstract}
\shortabstract
\begin{document}
\maketitle

\def\be{\begin{equation}}
\def\ee{\end{equation}}
\def\d {{\rm d}}
\def\del          {\partial}
\def\delbar       {\bar\partial}
\def\ii           {{\rm i}}
\def\tr           {\mathop{\rm Tr}}
\def\Re           {{\rm Re\hskip0.1em}}
\def\Im           {{\rm Im\hskip0.1em}}
\def\cala         {{\cal A}}
\def\calb         {{\cal B}}
\def\calc         {{\cal C}}
\def\cald         {{\cal D}}
\def\cale         {{\cal E}}
\def\calf         {{\cal F}}
\def\calg         {{\cal G}}
\def\calk         {{\cal K}}
\def\call         {{\cal L}}
\def\calm         {{\cal M}}
\def\caln         {{\cal N}}
\def\calo         {{\cal O}}
\def\calp         {{\cal P}}
\def\calr         {{\cal R}}
\def\cals         {{\cal S}}
\def\calt         {{\cal T}}
\def\calw         {{\cal W}}
\def\calz         {{\cal Z}}


\section{Introduction}
The highest possible rank of anti-symmetric gauge fields in four-dimensional space-time that allow for non-trivial fluxes is rank three. Such three-form gauge fields, i.e. $A_3$, will not carry propagating degrees of freedom in four dimensions because the on-shell values of their four-form field strengths $F_4={\rm d}A_3$ are constant. Nonetheless, the presence of these fields within a four-dimensional effective theory may have non-trivial physical implications. 
For instance, when coupled to gravity, four-form fluxes induce dynamically a contribution to the cosmological constant (see e.g. \cite{Duff:1980qv,Aurilia:1980xj,Hawking:1984hk,Brown:1987dd,Brown:1988kg,Duff:1989ah,Duncan:1989ug}). 
Moreover, when the three-form gauge fields couple to membranes, they provide a mechanism for dynamically neutralizing the cosmological constant via membrane nucleation \cite{Brown:1987dd,Brown:1988kg,Bousso:2000xa}. 
Apart from the aforementioned applications in late-time cosmology, gauge three-forms have been studied in inflationary models, e.g. in \cite{Bousso:2000xa,Kaloper:2008fb,Kaloper:2011jz,Marchesano:2014mla,Dudas:2014pva,Bielleman:2015ina,Valenzuela:2016yny}, 
and may play a role in the resolution of the strong CP violation problem, as discussed e.g. in  \cite{Dvali:2005an,Dvali:2004tma,Dvali:2005zk,Dvali:2013cpa,Dvali:2016uhn,Dvali:2016eay}. 

In four-dimensional ${\mathcal N}=1$ supersymmetric theories gauge three-forms can appear as part of variant matter and supergravity multiplets \cite{Gates:1980ay}. 
In such multiplets the Hodge-dual of the four-form field strength $*F_4$ will take the place of conventional scalar auxiliary fields. 
Supersymmetric models of variant three-form multiplets have been studied in a series of publications  \cite{Stelle:1978ye,Ogievetsky:1978mt,Ogievetsky:1980qp,Gates:1980ay,Gates:1980az,Buchbinder:1988tj,Binetruy:1996xw,Kuzenko:2005wh,Nishino:2009zz,Duff:2010vy,Groh:2012tf,Bandos:2016xyu,Aoki:2016rfz,Farakos:2017jme,Nitta:2018yzb,Nitta:2018vyc} 
and couplings with supersymmetric membranes have been also investigated 
\cite{Ovrut:1997ur,Huebscher:2009bp,Bandos:2010yy,Bandos:2011fw,Bandos:2012gz,Bandos:2018gjp}. 
These fields generically will arise from type-II string compactifications on Calabi--Yau manifolds with fluxes \cite{Bousso:2000xa,Bielleman:2015ina,Carta:2016ynn,Herraez:2018vae} and can thus provide the realization of the aforementioned physical scenarios in effective four-dimensional supergravity models. 
Interestingly, 
for variant matter three-form chiral multiplets, 
the non-zero expectation values of $*F_4$ fluxes may induce spontaneous supersymmetry breaking and generate a positive contribution to the cosmological constant. 
Such a setup can naturally incorporate {\it nilpotent} three-form chiral superfields \cite{Farakos:2016hly,Buchbinder:2017vnb}.

Generic chiral models with auxiliary fields of matter and supergravity multiplets traded for gauge three-forms were constructed in \cite{Farakos:2017jme}. This construction provided a manifestly supersymmetric effective description of type II  RR-flux compactifications with 4D three-forms that have been considered in \cite{Bielleman:2015ina,Carta:2016ynn,Herraez:2018vae}. 
The kappa-symmetric world-volume action for the supermembrane coupled to these systems has been constructed in \cite{Bandos:2018gjp}. Furthermore, in \cite{Bandos:2018gjp}, a general structure of BPS domain wall configurations sourced from the membranes was considered in detail. Such domain walls significantly enlarge the class of solutions previously described within standard matter-coupled supergravity \cite{Cvetic:1992bf,Cvetic:1992st,Cvetic:1992sf,Cvetic:1993xe,Cvetic:1996vr,Ceresole:2006iq} and in the three-form supergravity \cite{Ovrut:1997ur,Huebscher:2009bp}. Finally, the general models constructed in \cite{Farakos:2017jme,Bandos:2018gjp} may serve as a ground for the realization of the Brown--Teitelboim \cite{Brown:1987dd,Brown:1988kg} and the Bousso--Polchinski \cite{Bousso:2000xa} mechanisms for the dynamical neutralization of the cosmological constant within 4D supergravity theories. 

In this contribution, upon a brief discussion of the basic features of the Brown--Teitelboim and Bousso--Polchinski mechanisms, we will review the main results of \cite{Farakos:2017jme,Bandos:2018gjp}.


\section{Relaxing the cosmological constant by membrane nucleation}
\label{sec:BT}
One of the most challenging problems in modern theoretical physics is to uncover the origin of dark energy in our Universe. 
Assuming that dark energy is sourced from a cosmological constant will require a mechanism that can explain its extremely small value compared to the Planck scale $(\Lambda_{\rm obs} \sim 10^{-122} M_P^4)$. 
Brown and Teitelboim in \cite{Brown:1987dd,Brown:1988kg} proposed a mechanism that 
relaxes the cosmological constant dynamically by a gauge three-form field nucleating membrane bubbles. 
Within such setup, 
an initial arbitrarily large cosmological constant 
can be reduced to a very small value 
due to subsequent membrane nucleations. 
This procedure reduces the value of the cosmological constant to lie within the so-called ``Weinberg's window'' \cite{Weinberg:1987dv}. 
However, as we will demonstrate later on, in this simple scenario the membranes should be heavy and {\it unnaturally} weakly coupled to the three-form gauge field. 

Let us summarize how the Brown--Teitelboim mechanism works. 
It comprises a membrane coupled to a single real gauge three-form $A_3$ and four-dimensional gravitation. 
Let us denote with $A_{mnp}$ the components of $A_3$ and with $F_{mnpq} = 4 \del_{[m} A_{npq]}$ those of 
$F_4$.
The  membrane world-volume $\calc$ is parametrized by the coordinates $\xi^\mu$ ($\mu=0,1,2$). 
The membrane dynamics are captured by the embedding of $\calc$ 
into the ambient space-time described by the coordinates $x^m$ ($m=0,1,2,3$), 
namely 
\be\label{xi-xxi}
\xi^\mu \quad\mapsto\quad  x^m (\xi)\,.
\ee
The action describing the interactions among gravity, 
the gauge three-form and the membrane is 
\be
\label{BTaction}
\begin{split}
	S &=  -\int \d^4x\, e\left( \frac{R}{2} + \lambda \right) + \int_{\calb} \d^3x\sqrt{-h} K\,+
	\\
	&\quad+\,\frac{1}{2} \int \d^4x\, e ({}^*\!F_4)^2 + \frac{1}{3!} \int \d^4x\,e\, \nabla_m \left(A_{npq} F^{mnpq}\right)-
	\\
	&\quad-\,T_{\rm M} \int_{\calc} \d^3 \xi\, \sqrt{-h}+\frac{q}{3!} \int_{\calc} \d^3 \xi\, \varepsilon^{\mu\nu\rho} \frac{\del x^m}{\del \xi^\mu}\frac{\del x^n}{\del \xi^\nu}\frac{\del x^p}{\del \xi^\rho}\,A_{mnp}\,.
\end{split}
\ee
The gravitational sector of the first line of \eqref{BTaction} contains the scalar curvature, a ``bare'' cosmological constant and 
the Gibbons--Hawking boundary term, 
where $K$ is the extrinsic curvature and $\calb$ denotes the boundary. 

The second line contains the kinetic term for the gauge three-form, expressed in terms of the Hodge-dual 
of its field strength ${}^*\!F_4 = -\frac{e}{4!} \varepsilon_{mnpq}F^{mnpq}$ 
(where $\epsilon^{0123}=$ $-\epsilon_{0123}=1$ and $e=\sqrt{-\det g_{mn}}=\det e^a_m$). 
The additional boundary term for the gauge three-forms ensures that the variation of the action 
with respect to the gauge three-form is compatible with proper gauge invariant boundary conditions, 
that are $\delta F_{mnpq}|_{\rm bd} = 0$, 
rather than the gauge variant $\delta A_{mnp}|_{\rm bd} = 0$ (see e.g. \cite{Brown:1987dd,Brown:1988kg,Groh:2012tf}). 

Finally, the third line of \eqref{BTaction} describes the dynamics of the membrane. 
The first term is the Nambu--Goto action with $T_{\rm M}$ being the membrane tension 
and $h_{\mu \nu}$ being the world-volume pullback of the space-time metric $g_{mn}$, or {\it induced metric},  
\be
h_{\mu\nu}(\xi)\equiv \frac{\del x^m}{\del \xi^\mu} \frac{\del x^n}{\del \xi^\nu} g_{mn} (x(\xi))\,, \qquad h\equiv \det h_{\mu\nu}(\xi).
\ee
Indeed, the last term in the third line describes the minimal coupling of the membrane of charge $q$ 
to the pull-back of the gauge three-form. 
In a more compact 
notation, this term can be written as $-q \int_{{\cal C}} A_3$.

From \eqref{BTaction}, 
we can derive the equation of motion for the gauge three-form $A_{mnp}$, 
which is given by 
\be\label{eomA}
\begin{split}
&\del_m {}^*\! F_4=  
\\
&=\frac{q}{3!} \int \d^3 \xi\, \delta(x-x(\xi))\,\varepsilon_{mnpq}\, \varepsilon^{\mu\nu\rho} \frac{\del x^n(\xi)}{\del \xi^\mu}\frac{\del x^p(\xi)}{\del \xi^\nu}\frac{\del x^q(\xi)}{\del \xi^\rho}\,. 
\end{split}
\ee
Away from the membrane, 
this equation is simply solved by ${}^*\! F_4 = E$, with $E$ being a real constant. 
Plugging this value back into the action \eqref{BTaction} 
and taking into account the crucial contribution from the boundary terms, one gets
a cosmological constant of the form $\Lambda = \lambda + E^2/2$. 
We thus observe that the cosmological constant term receives a \emph{dynamical contribution} from the gauge three-form. In other words, setting the gauge three-form on-shell alters the original bare value $\lambda$ of the cosmological constant. 

Let us now see how the presence of the membrane influences the \emph{dynamically generated cosmological constant}. If the membrane surface is closed, 
it will divide the ambient space-time into an \emph{outside} region and an \emph{inside} region. 
The constant value of the four-form flux $E$ will be different in the two regions, 
and the difference between the two values can be readily computed from \eqref{eomA}. 
Let us call $E_{\rm I}$ the constant flux inside the membrane and $E_{\rm O}$ the one outside. 
There is a coordinate system which allows us to effortlessly relate these values with the membrane charge $q$: 
the so-called {\it static gauge}. 
We adopt a local coordinate system (in the vicinity of a point on the membrane) 
such that three of the space-time coordinates coincide with the world-volume coordinates, 
namely we set $x^\mu=\xi^\mu$. 
The world-volume is then described by the equation $x^3=x^3(\xi)=0$\footnote{Such a gauge can be fixed by target-space general 
coordinate transformations which is a gauge symmetry of our dynamical system \eqref{BTaction}. 
This reflects the  Goldstone nature of the membrane coordinate functions, 
which transform as St\"uckelberg fields in the presence of dynamical gravity. 
See \cite{Bandos:2001jx} for a discussion and references on this issue.}. 
Now, in this neighbourhood, the $m=3$ component of \eqref{eomA} reads
\be\label{d3*F=}
\del_3 {}^*\! F_4 = q \, \delta(x^3) \int \d^3 \xi\,  \delta^3 (x^\mu- \xi^\mu)=   q \, \delta(x^3)\; .
\ee
Once equation \eqref{d3*F=} is integrated along $x^3$ over a small (infinitesimal) interval, 
say from $-\epsilon$ to $+\epsilon$, one gets
\be 
E_{\rm O} - E_{\rm I} = q \, . 
\ee
Hence, 
the value of the cosmological constant in the outside region compared to its value in the inside region changes as 
\be
\label{RedCC}
\Lambda_{\rm O} =  \lambda+ \frac{E_{\rm O}^2}{2} \;\rightarrow\; \Lambda_{\rm I} = \lambda+\frac{E_{\rm I}^2}{2} =  \lambda+\frac{(E_{\rm O}-q)^2}{2} \, . 
\ee

If we assume that $E_{\rm O}>0$ and $q>0$, 
then the membrane encloses a region with a lower value of the cosmological constant. 
As a result, membrane solutions can be considered as domain wall type solutions 
that separate two configurations specified by the fluxes $E_{\rm O}$ and $E_{\rm I}$. 
The membrane nucleation procedure is determined by quantum tunnelling processes described by instantons \cite{Brown:1987dd,Brown:1988kg}. 
In concrete models, the bare contribution $\lambda$ to the cosmological constant
is (usually) negative (eventually, it can be of $\calo (1)$ in Planck units) 
and $E_{\rm O}= n q$ is generically an integer multiple of the charge of the membrane. 

The neutralization of the cosmological constant proceeds as follows. Let us assume that the space-time, filled with a three-form field, sits in a vacuum characterized by the cosmological constant $\Lambda_{\rm O} =  \lambda+ \frac{E_{\rm O}^2}{2}$. Let us assume that $\frac{E_{\rm O}^2}{2} > |\lambda|$, so that the starting vacuum is a dS space-time. At a certain instant, a membrane is created as a soliton which allows for the decay of the vacuum. The creation of such an object carries away an amount of energy which depends on the charge of the membrane. Once the membrane is created, the space is divided into an outside region, still with cosmological constant $\Lambda_{\rm O}$, and an inside region with a lower value $\Lambda_{\rm I}$. However, this is not the end of the procedure. The internal vacuum with $\Lambda_{\rm I}$ may further decay to a vacuum with an even lower value of the cosmological constant, and so on. The subsequent nucleation of membranes, which is accompanied with a decrease of the cosmological constant, ceases when the cosmological constant is almost neutralized. 
This happens when the cosmological constant gets negative and in addition satisfies 
\be
- |q E_{\rm final}| < \Lambda_{\rm final} <0\,.
\ee
When such a condition is met, an instanton solution does not exist and the membrane nucleation stops \cite{Brown:1987dd,Brown:1988kg}. 
Alternatively, for a de Sitter vacuum, one finds it can have a large life-time depending on how small the membrane charge is. 
As a result, this mechanism could have explained why the cosmological constant is very close to zero. 
However, some issues arise that make the model unrealistic, as noticed readily in the original works \cite{Brown:1987dd,Brown:1988kg}. 

First, there is the the \emph{step size problem}. 
Assume that $n \simeq \sqrt{2 |\lambda|}/q$ is the number of nucleations such that the cosmological constant is almost neutralized. 
We might expect, in order to avoid fine-tuning, that at each step of membrane nucleation, 
the cosmological constant should be decreased by an amount at most of order $\Lambda_{\rm obs}$. This implies  $|q| \sim |\lambda|^{-\frac12} \Lambda_{\rm obs} $. 
The value of $|\lambda|$ can be of order one (but generically not smaller than the lower bound 
on the supersymmetry breaking scale $|\lambda| \gtrsim 10^{-64}$ \cite{Bousso:2007gp}). 
Consequently, the charges of the membrane should be unnaturally small $|q| \lesssim 10^{-90}$. 
Clearly in this scenario one trades the problem of explaining the smallness of the cosmological constant 
with the problem of explaining the smallness of the membrane charge. 

The second relevant issue is the so-called \emph{empty universe problem}. 
As found in \cite{Brown:1987dd,Brown:1988kg}, as a consequence of the tiny membrane charges,  membrane nucleations happen with time intervals of order $\sim \exp({10^{120}})$sec. 
Therefore, the universe would experience a so prolonged de Sitter expansion between the nucleations of membranes 
such that all matter would be completely diluted before reaching $\Lambda_{\rm obs}$.

In \cite{Bousso:2000xa}, 
Bousso and Polchinski 
noted that both these problems could be resolved in the framework of string/M-theory compactifications in which systems similar to those described by the action
 \eqref{BTaction} are produced naturally. Indeed, on the one hand, dimensional reduction of p-form gauge fields of the 11/10D theories to 4D 
produces a plethora of gauge three-forms $A_{3I}$  ($I=1,\ldots,N$, with $N$ depending on the geometry of the compactifying manifold) and their associated four-form fluxes are quantized. On the other hand, membranes may originate from the dimensional reduction of higher dimensional branes wrapped on some internal cycles and they naturally couple, with generic charges $q_I$, to the various gauge three-forms $A_{3I}$.

After a single membrane nucleation the cosmological constant changes as follows 
\be
 \label{RedCCBP}
 \Lambda_{\rm O} = \lambda+ \frac12 \sum\limits_{I=1}^Nn^2_I q^2_I \;\rightarrow\; \Lambda_{\rm I} =\lambda + \frac12 \sum\limits_{i=1}^N (n_I-1)^2 q^2_I \, , 
 \ee
where $n_I$ are units of flux quanta. 
Because of the dependence on multiple $q_I$, the units of fluxes $n_I$ form a dense \emph{discretum}, allowing for values of the cosmological constant which can be well below $\Lambda_{\rm obs}$, 
without assuming that the charges are unnaturally small 
and with a bare cosmological constant $\lambda$ of order $\calo(1)$. 
For example, $\Lambda_{\rm obs}$ may be achieved for $N \sim 100$ with $|\lambda| \sim \calo(1)$ and $|q| \sim \calo(10^{-1})$.

The empty universe problem can now be easily bypassed.
Since the step of each nucleation is not required to be small, it is sufficient to assume that, before the last nucleation, the value of the cosmological constant was much larger than $\Lambda_{\rm obs}$. 
Assuming then the existence of an inflaton, 
\emph{before} the final membrane nucleation 
the quantum fluctuations will dominate over its classical motion driving it far from the vacuum. 
In some regions of space, 
the inflaton will be at appropriate values such that 
\emph{after} the final nucleation, 
inflation will start because the classical motion of the inflaton will dominate. 
Our universe will emerge from one of these small regions where inflation takes place. 
Then, the empty universe problem is solved in the same way as it is solved in standard inflationary cosmology, 
namely after the exit from inflation a {\it reheating} phase will repopulate the universe.

In summary, 
the effective theories arising from string compactifications seem to provide a 
solid  framework to implement the aforementioned mechanisms. 
Clearly, the Bousso--Polchinski proposal crucially relies on the existence of multiple gauge three-forms and of the corresponding membranes, which are indeed expected for generic string compactifications. Still, a model like \eqref{BTaction}, even if extended to include multiple three-forms and membranes, cannot be the final answer. In typical compactifications, plenty of scalar moduli are generically present, which non-trivially interact with the three-form fluxes. We should then expect that also these scalar fields experience some \emph{jumps} across the membranes, exactly like the cosmological constant. The full solitonic solutions should necessarily take into account the possible interactions among gauge three-forms, membranes and moduli.

In what follows, we will briefly review the results of \cite{Farakos:2017jme,Bandos:2018gjp,Farakos:2017ocw} 
on the description of these effective theories within a generic 4D 
${\mathcal N}=1$ supergravity containing multiple three-form gauge fields coupled to membranes.


\section{Gauge three-form supergravity}
\label{sec:3formSugra}

In \cite{Farakos:2017jme}, there was proposed a general receipt of how to incorporate gauge three-forms into 4D supergravity theory by dualizing chiral multiplets into so-called \emph{double} (and \emph{single}) \emph{three-form} multiplets. 
Simple examples have been constructed in \cite{Stelle:1978ye,Ogievetsky:1978mt,Ogievetsky:1980qp,Gates:1980ay,Gates:1980az,Buchbinder:1988tj} 
and further developments can be found in 
\cite{Binetruy:1996xw,Kuzenko:2005wh,Nishino:2009zz,Duff:2010vy,Groh:2012tf,Farakos:2016hly,Bandos:2016xyu,Aoki:2016rfz,Farakos:2017jme,Nitta:2018yzb,Nitta:2018vyc,Cribiori:2018jjh}. 
These multiplets are also chiral, but they differ from the ordinary ones in their highest components. Rather than being auxiliary complex scalar fields, they are particular combinations of real gauge three-forms. 

Let us consider a generic matter-coupled $\caln=1$ (minimal) supergravity whose matter content consists of two distinct sets of chiral 
superfields: 
$T^r$, with $r=1,\ldots,m$, which are ordinary chiral  
superfields  describing scalar multiplets with zero scaling dimension $\Delta_T=0$, and 
special chiral superfields
$S^I$, with $I=0,\ldots,n$, which have scaling dimension $\Delta_S=3$ and 
describe double three-form multiplets. We also assume that the manifolds parametrized by $S^I$ and $T^r$ factorize and that $S^I$ 
are coordinates of a special K\"ahler manifold locally specified by a prepotential $\calg(S)$ which is holomorphic in $S^I$ and homogeneous of order two  
\be\label{cG=hom}\calg(\mu S)=\mu^2 \calg(S)\; . \ee
In the following we will denote 
\be\label{gs}
\calg_I(S)\equiv\del_I\calg(S)=\calg_{IJ}S^J$, $\calg_{IJ}\equiv \del_I\del_J\calg(S).
\ee
More explicitly, the double three-form superfields $S^I$ are defined by the following non-linear relations
\begin{subequations}
\be\label{Sdef}
S^I\equiv \frac14(\bar\cald^2-8\calr)\calm^{IJ}(S)(\Sigma_J-\bar\Sigma_J)\, , 
\ee
where 
\be
\label{cM:=}\calm^{IJ} \equiv (\calm_{IJ})^{-1}\,,  \qquad \calm_{IJ}\equiv \Im\calg_{IJ}\; , 
\ee 
and $\Sigma_I$ are complex linear superfields obeying the constraint 
\be
\label{bD2Si=0}(\bar\cald^2-8\calr)\Sigma_I=0 \, . 
\ee 
\end{subequations}
The superfields $\Sigma^I$ accommodate two sets of real gauge three-forms $A_3^I$ and $\tilde A_{3I}$ as their components 
\be\label{compSigma}
\begin{aligned}
    s^I&=S^I\big|_{\theta=\bar\theta=0} \, , 
    \\
	\bar\sigma_m^{\dot\alpha\alpha}[\cald_\alpha,\bar\cald_{\dot\alpha}]\Sigma_I\big|_{\theta=\bar\theta=0}&=-2\big({}^*\!\tilde A_{3I}-\calg_{IJ}(s)\, {}^*\!A_3^J\big)_m \,.
\end{aligned}
\ee
The highest components of $S^I$ contain the Hodge-duals of their four-form field strengths
\be\label{SF}
F^I_S\equiv -\frac14\cald^2 S^I|_{\theta=\bar\theta=0}=\bar M s^I-\frac\ii2\,\calm^{IJ}{}^*\!\calf_{4J}\,,
\ee
where $M$ is the complex scalar auxiliary field of the supergravity multiplet and we have defined
\be\label{defcalf}
\calf_{4I}\equiv \tilde F_{4I}-\bar\calg_{IJ} F_4^J\,.
\ee
These are complex four-forms, depending on the field strengths of the real three-forms $ F_4^I \equiv \d A_3^I$ and $ \tilde F_{4I} \equiv \d \tilde A_{3I}$. 

The most general, super-Weyl invariant Lagrangian which can be built with no superpotential is
\be\label{3formlagr}
\call_{\rm SG}=-3\int\d^4\theta\,E\,\Omega(S,\bar S;T,\bar T)+ \text{c.c.} +\call_{\rm bd}\,.
\ee
Here, as in \eqref{BTaction}, owing to the presence of gauge three-forms, boundary terms should be included to ensure the correct variation of the Lagrangian and $\Omega(S,\bar S;T,\bar T)$ is the kinetic function which has scaling dimension $\Delta_\Omega=2$. In our setup, $\Omega$ factorizes as
\be\label{Omega}
\begin{split}
\Omega(S,\bar S;T,\bar T)&=\Omega_0(S,\bar S) e^{-\frac13 \hat K(T,\bar T)} \, ,\\ 
\Omega_0(S,\bar S) &=\left[\ii \bar S^I\calg_I(S)-\ii S^I\bar\calg_I(\bar S) \right]^\frac{1}{3} \, .
\end{split}
\ee
In order to compute the bosonic components of the Lagrangian \eqref{3formlagr}, we first have to fix the super-Weyl invariance. To this aim, we write $S^I$ in terms of a chiral compensator $Y$, carrying scaling dimension $\Delta_Y=3$ and the `physical' chiral superfields $\Phi^i$, with $i=1,\ldots,n$, having $\Delta_\Phi=0$. 
We set 
\be\label{SYPHI}
S^I=Y f^I(\Phi) \, , 
\ee
where $f^I(\Phi)$ are holomorphic functions of $\Phi^i$ such that $\text{rank}(\del_if^I)=n$. 

We may now gauge fix super-Weyl symmetry by setting $Y=1$, perform 
the integration over the fermionic variables  and 
a Weyl rescaling to pass to the Einstein frame and, after having integrated out the complex auxiliary fields $F^r_T$ residing inside the multiplets $T^r$, we arrive at the bosonic action 
\be\label{boscomp}
\begin{aligned}
	S_{\rm SG,\, bos}&=- \int \d^4 x\, e \Big(\frac{1}{2}R + \; \del \phi^i \del \bar{\phi}^{\bar \jmath}  + \hat K_{p \bar q}\,\del t^p \del\bar t^{\bar q}\,-
	\\
	&\kern2.5cm-\,\calt^{IJ}{}^*{}\!\bar\calf_{4I} {}^*{}\!\calf_{4J}\Big)+S_{\rm bd} \,, 
\end{aligned}
\ee
with the boundary terms
\be\label{bdaction}
\begin{aligned}
	S_{\rm bd} 
	&=-2\Re\int \d^4 x\, e\,\nabla_m \left[ \calt^{IJ}\big( {}^*\!\tilde A_{3I}-\calg_{IK}{}^*\!A_3^K\big)^m\,{}^*\!\calf_{4J}\right]\,.
\end{aligned}
\ee
In \eqref{bdaction} the space-time boundary is understood to be at infinity. The quantities appearing in \eqref{boscomp} are all given in terms of the prepotential $\calg(S)$ and the K\"ahler potential $\hat K (T, \bar T)$ as
\begin{subequations}
	\begin{align}
	K_{i\bar\jmath}&\equiv G_{IJ} f^I{}_i \bar{f}^{J}{}_{\bar\jmath}\,,
	\\
	G_{IJ}&\equiv-\frac{\calm_{IJ}}{(f \calm \bar f)}+\frac{(\calm\bar f)_I(\calm f)_{J} }{(f \calm \bar f)^2},\label{GIJ}\\ 
	\calt^{IJ}&\equiv \frac14\, e^{-\calk}\left[\calm^{IK}G_{LK}\calm^{LJ}+\frac{1}{\gamma}\frac{f^I\bar f^J}{(f\calm\bar f)^2}\right]\,,\\
	\gamma &\equiv \hat{K}_{\bar q} \hat{K}^{\bar q p} \hat{K}_{p}-3\label{noscale}\,,
	\end{align}
\end{subequations}
with $(\calm f)_I\equiv \calm_{IJ}f^J$, $(\calm \bar f)_I\equiv \calm_{IJ}\bar f^J$ and $( f \calm \bar f)\equiv f^I \calm_{IJ}\bar f^J$. 

As in Section \ref{sec:BT}, we can now obtain the potential by setting the gauge three-forms on-shell. By solving their equations of motion, we get
\be\label{eom3f}
2\Re(\calt^{IJ}{}^*\!\calf_{4J})=m^I\,,\qquad 2\Re(\calg_{IJ}\calt^{JK}{}^*\!\calf_{4K})=e_I,
\ee
where $e_I$ and $m^I$ are real quantized constants. The Lagrangian \eqref{3formlagr} then becomes
\be\label{boscompOS}
\begin{aligned}
	S_{\rm bos}=&- \int \d^4 x\, e \Big(\frac{1}{2}R + G_{IJ} f^I{}_i \bar{f}^{J}{}_{\bar\jmath}\; \del \phi^i \del \bar{\phi}^{\bar \jmath}\,+
	\\
	&\kern2cm+\,\hat K_{p \bar q}\,\del t^p \del\bar t^{\bar q}+V(\phi,\bar\phi,t,\bar t;e,m)\Big)\,,
\end{aligned}
\ee
where the potential $V(\phi,\bar\phi,t,\bar t;e,m)$ for the scalar fields $\phi^i$ and $t^q$ is given by
\be\label{potst}
\begin{aligned}
	V&=\calt^{IJ}{}^*\!\bar\calf_{4I}\, {}^*\!\calf_{4J}|_{\text{on-shell}} \, . 
\end{aligned}
\ee
The same form of the potential can be calculated from the standard supergravity equation  
\be\label{potstW}
\begin{aligned}
	V&=e^\calk\left( K^{i\bar\jmath}D_iW\bar D_{\bar\jmath}\overline W+\gamma|W|^2\right) ,
\end{aligned}
\ee
with 
the superpotential 
\be
\label{superp}
W=  e_If^I(\phi)-m^I\calg_I(\phi)\,.
\ee
I.e. the potential \eqref{potst} is the same as the one obtained from an ordinary chiral matter supergravity model with superpotential $W$. The  
role of the gauge three-forms has been to dynamically generate the parameters $e_I$ and $m^I$ appearing in the superpotential, 
therefore promoting them to vacuum expectation values of the field strengths of the gauge three-forms. 


\section{Supergravity coupled to membranes}
\label{sec:Membranes}

To reproduce an action like \eqref{BTaction} within 4D supergravity, one should couple membranes to three-forms described by \eqref{3formlagr} in such a way that the  local supersymmetry is preserved. As was considered e.g. in \cite{Achucarro:1988qb,DeAzcarraga:1989vh,Ovrut:1997ur,Huebscher:2009bp,Bandos:2010yy,Bandos:2011fw,Bandos:2012gz,Kuzenko:2017vil}, membranes can indeed be promoted to objects moving in the whole four-dimensional superspace. 
In particular, a kappa-symmetric world-volume action describing the coupling of a supermembrane to a variant of minimal ${\mathcal N}=1$ supergravity with a single gauge three-form field  was considered in \cite{Ovrut:1997ur,Bandos:2011fw,Bandos:2012gz} and a kappa-symmetric action for a supermembrane coupled to a single three-form chiral matter superfield in flat ${\mathcal N}=1$ superspace was constructed in \cite{Bandos:2010yy}. 

A space-time component action describing the interaction of a membrane with supergravity and scalar matter multiplets, in which the membrane part of the action is purely bosonic, was considered in \cite{Huebscher:2009bp}. Conceptually, this formulation is similar to the `complete but gauge-fixed description' of dynamical systems of supergravity and super-p-branes whose general concept was introduced and considered on various examples in \cite{Bandos:2001jx,Bandos:2002kk,Bandos:2002bx,Bandos:2003zk,Bandos:2005ww}.

The coupling of the supermembrane to double three-form supergravity in its super-Weyl invariant formulation was recently constructed in \cite{Kuzenko:2017vil}. This reference also gives a detailed review of different variants of the old minimal supergravity. For a discussion of top-form supergravities and their coupling to strings in three-dimensional space-time see \cite{Buchbinder:2017qls}.

The embedding of the membrane world-volume, parameterized by $\xi^\mu$, into the target superspace ($z^M$), 
is given by 
\be\label{superembedding}
\xi^\mu\quad\mapsto\quad \calc:\;z^M(\xi)=\left(x^m(\xi),\theta^\alpha(\xi),\bar\theta^{\dot\alpha}(\xi)\right). 
\ee
The membrane dynamics is governed by the superspace action  
\be\label{SM}
S_{\rm M}\equiv S_{\rm NG}+  S_{\rm WZ}\,.
\ee

The first part of this action, that is the Nambu--Goto term, reads 
\be\label{SmembrNG} S_{NG}= -2\int_{\calc} \d^3\xi \sqrt{- h} \; | T| \; , \qquad
\ee
where $h=\det h_{\mu\nu}$ is the determinant of the induced metric
\be\label{hmunu}
h_{\mu\nu}(\xi)\equiv \eta_{ab}E^a_\mu(\xi)E^b_\nu(\xi)
\ee
constructed from the pull--backs of the target superspace supervielbein
\be
E^a_\mu(\xi)\equiv  \del_\mu z^M(\xi) E^a_M(z(\xi)) \, .
\ee
The supervielbein forms $E^A=dZ^M E_M^A(Z)=(E^a, E^\alpha, E^{\dot{\alpha}})$ satisfy the minimal supergravity constraints, 
the explicit form of which can be found in  \cite{Wess:1992cp}. 
Finally,  $T$ is the pull--back of a (fundamental or composite) covariantly chiral superfield obeying 
\be\label{bDT=0} \bar{{\cal D}}_{\dot{\alpha}}T=0 \; . \ee
Let us observe that $|T|$ plays the role of an effective tension of the supermembrane. 

The second part of the action \eqref{SM}, that is the Wess--Zumino term, has the form 
\be\label{SmembrWZ} 
S_{WZ}=  \int_{\calc}  {\cal A}_3 
\ee
and describes the coupling of the membrane to the pull-back of a three-superform  ${\cal A}_3$. 
This term is the supersymmetrization of the  minimal coupling appearing in \eqref{BTaction}. 
The local fermionic $\kappa$--symmetry of the supermembrane action, which we will discuss below, 
requires that ${\cal A}_3$ and the chiral superfield $T$ 
are related to each other. In particular, the four-form field strength of ${\cal A}_3$ is expressed in terms of $T$ as follows 
\begin{eqnarray} \label{bH4=SGZ}   {\cal H}_{4}   &=& \d{\cal A}_3= \;   {i} E^b\wedge E^a \wedge \bar E^{\dot\alpha} \wedge \bar  E^{\dot\beta }\bar{\sigma}_{ab\; \dot{\alpha}\dot{\beta}} T\,- \qquad \nonumber \\  
&& - {i} E^b\wedge E^a \wedge E^\alpha \wedge E^\beta \sigma_{ab\; \alpha\beta}\bar{ T}\,-  \qquad \nonumber \\  
&& - \frac{i}6  E^c\wedge E^b\wedge E^a \wedge \bar E^{\dot\alpha} \epsilon_{abcd} \sigma^d_{\alpha\dot\alpha} {\cal D}^{\alpha}{ T}\,-\qquad \nonumber \\  
&& -\frac{i}6 E^c\wedge E^b\wedge E^a \wedge E^\alpha \epsilon_{abcd} \sigma^d_{\alpha\dot\alpha} \bar{{\cal D}}^{\dot\alpha}\bar{ T}\,+ \nonumber \\  
&& + \frac{1}{96} E^{d} \wedge E^c \wedge E^b \wedge E^a \epsilon_{abcd} \left(({\cal D}{\cal D}-24\bar{{{\cal R}}})
{ T}+ \right. \qquad \nonumber \\  
&& \kern1cm\left.\qquad  +(\bar{{\cal D}}\bar{{\cal D}}-24{{\cal R}})
\bar{ T} \right)
\,,\qquad
\end{eqnarray}
where $\mathcal R$ is a chiral superfield of old minimal supergravity \cite{Wess:1992cp}. 

For the four-form \eqref{bH4=SGZ} to be exact, one requires that the chiral superfield $T$ is `special', in the sense that it is constructed in terms of a real rather than a generic complex pre-potential superfield ${\cal P}$. 
For $T$ we have 
\be\label{cZqp=bD2P}  T = -\frac i4 (\bar{{{{\cal D}}}}^2-8 {{\cal R}}) \cal P,  \qquad {\cal P}=({\cal P})^*\; . \qquad 
\ee
Then (modulo gauge transformations) the three-form potential is constructed with the use of $\mathcal P$ as follows
\pagebreak
\be\label{super3form}
\begin{aligned}
{\cal A}_{3}=&\,  { -}2 {i} E^a \wedge E^\alpha \wedge \bar E^{\dot\alpha}  \sigma_{a\alpha\dot\alpha}{{\cal P}}\,+ \\
& +\, {\frac 12}  E^b\wedge E^a \wedge  E^\alpha
\sigma_{ab\; \alpha}{}^{\beta}{{{\cal D}}}_{\beta}{{\cal P}}\,+ \\ 
&+\,{\frac 12}  E^b\wedge E^a \wedge  \bar E^{\dot\alpha}\bar\sigma_{ab}{}^{\dot\beta}{}_{\dot\alpha}\bar{{{\cal D}}}_{\dot\beta}{{\cal P}}\,+
\\
&+\,\frac {1} {24} E^c \wedge E^b \wedge E^a \epsilon_{abcd} \,\left(\bar{\sigma}{}^{d\dot{\alpha}\alpha}
  [{{\cal D}}_\alpha, \bar{{{\cal D}}}_{\dot\alpha}]{{\cal P}}-3 G^d{{\cal P}} \right)
 \, ,
\end{aligned}
\ee
where $G^d$ is the real vector superfield of minimal supergravity 
satisfying $\overline{\cal D}^{\dot \alpha} G_{\alpha \dot \alpha}= {\cal D}_{\alpha} {\cal R}$ \cite{Wess:1992cp}. 

When $T$ is identified with the chiral compensator of a Weyl transformation of the superconformally invariant formulation of supergravity, the above action describes the supermembrane interacting with three-form variants of pure minimal supergravity \cite{Kuzenko:2017vil}. In particular, if $\mathcal P$ is a generic real scalar superfield one deals with  
the single three-form  supergravity  and arrives at the action of \cite{Ovrut:1997ur,Bandos:2012gz} upon gauge fixing $T=1$. While, if $\mathcal P$ is the real part of a complex linear superfield the membrane is coupled to the double three-form supergravity \footnote{This formulation can also be  extracted from the equations in \cite{Bandos:2012gz}, although there it was not described in detail.}. When $T$ is regarded as an independent matter superfield, the complete superfield theory action should also include the kinetic term and, possibly, superpotential for this chiral superfield. In this case the supergravity multiplet may be the standard old-minimal one. In flat superspace the action for such an interacting system was studied in \cite{Bandos:2010yy}. 

Here, as in \cite{Bandos:2018gjp}, our main interest is a generic coupling of supermembranes to double three-form matter and supergravity multiplets.
In this case  $ T$ is a composite special chiral superfield
\be\label{cZqp}  T = q_I S^I- p^I {\cal G}_I(S)=
q_IS^I -p^I{\cal G}_{IJ}(S)S^J \, ,
\ee
where $S^I$ and ${\cal G}(S)$ were defined in (\ref{Sdef}) and \eqref{gs}, and $q_I$ and $p^I$ have, as we will see below, the meaning of `electric' and `magnetic' charges of the supermembrane with respect to various three-form potentials. 

From (\ref{Sdef}) one can see that the chiral superfield 
(\ref{cZqp})  is constructed as in (\ref{cZqp=bD2P})
from the composite prepotential
\be\label{WZprepot1} \begin{aligned}
     {{\cal P}} & = q_I {{\cal P}}^I- p^I \tilde{{\cal P}}_I 
\end{aligned}
\ee
where 
\be\label{realprepot1}
{{\cal P}}^I\equiv -2 {{\cal M}}^{IJ}{\rm Im}\Sigma_J\quad,\quad \tilde{{\cal P}}_I\equiv -2{\rm Im}(\bar{\cal G}_{IJ}{{\cal M}}^{JK}\Sigma_K) \, ,
\ee
and $\Sigma_I$ are complex linear superfields defined by 
(\ref{bD2Si=0}).  
We then find that
\be\label{defSG}
S^I=-\frac i4(\bar{{\cal D}}^2-8{{\cal R}}){{\cal P}}^I\,,\quad {\cal G}_I=-\frac i4(\bar{{\cal D}}^2-8{{\cal R}})\tilde{{\cal P}}_I \, .
\ee

Introducing the 3--form potentials ${\cal A}_3^I$ and $\tilde{{\cal A}}_{3I}$ constructed as in (\ref{super3form}) with the composite prepotentials (\ref{realprepot1}), we find that the integrand of the Wess--Zumino term (\ref{SmembrWZ}) 
is 
\be\label{calA3=A+tA}
\cala_{3}= q_I \cala_{3}^I - p^I \tilde{\cala}_{3I} . 
\ee
This makes it clear that the constants $q_I$ and $p^I$ play the role of the electric and magnetic charges of 
the membrane coupled to the three-form gauge fields. 

To resume, the action for the membrane interacting with double-three form matter and supergravity is given by (\ref{SM}) where 
the Wess--Zumino term is
\be
\label{susyWZ}
S_{\rm WZ}=\int_{{\cal C}}(q_I \cala^{3I} - p^I \tilde{\cal A}_{3I})
\ee
and the Nambu--Goto term is
\be\label{susyNG}
S_{\rm NG}=-2\int_\calc\d^3\xi\sqrt{- h}\left|q_IS^I-p^I\calg_I(S)\right|.
\ee

Let us now discuss the symmetry properties of the action \eqref{SM}, \eqref{susyWZ} and \eqref{susyNG}.
\begin{description} 
	\item[world-volume reparameterization invariance.] The action\linebreak \eqref{SM} is clearly invariant under reparametrizations of the membrane world-volume $\xi \rightarrow \xi'(\xi)$. A convenient way to fix this freedom is to set
	\be
	\label{staticg}
	\xi^\mu \equiv x^\mu\,.
	\ee
	This gauge choice leaves the fourth coordinate $x^3\equiv y(\xi)$ as the only bosonic physical field describing the dynamics of the membrane. 
	The field $y(\xi)$ can then be interpreted as the transverse displacement of the membrane from its static position. However, the static configuration breaks the translation invariance of the background, and $y(\xi)$ plays the role of the Goldstone field associated with this spontaneous breaking. 
	\item[$\kappa$-symmetry.] The action \eqref{SM} also enjoys a peculiar local fermionic symmetry, called \emph{$\kappa$-symmetry}, which acts on the space-time coordinates as follows
	\be\label{kappasymm}
	\delta z^M (\xi)=\kappa^\alpha(\xi) E^M_\alpha(z(\xi))+\bar\kappa^{\dot\alpha}(\xi) E^M_{\dot\alpha}(z(\xi))\,. 
	\ee
	Here $\kappa^\alpha(\xi)$ (with $\bar\kappa^{\dot\alpha}(\xi)\equiv 	\overline{\kappa^\alpha}(\xi)$) is a local fermionic parameter satisfying the projection condition 
	\be\label{kappaproj}
	\kappa_\alpha=\frac 	{q_IS^I-p^I\calg_I}{|q_IS^I-p^I\calg_I|}\Gamma_{\alpha\dot\alpha}\bar\kappa^{\dot\alpha},
	\ee
	with \footnote{In the generic case of (\ref{SmembrNG}), (\ref{SmembrWZ}) and (\ref{bH4=SGZ}), the restriction on 
	the $\kappa$-symmetry parameter reads 
	$
\kappa_\alpha =\frac T {|T|} ({\Gamma}\bar{\kappa})_\alpha$ where $\Gamma_{\alpha\dot\alpha}$ 
is given by  (\ref{bGamma}).
}
	\be\label{bGamma}
	\Gamma_{\alpha\dot\alpha}\equiv \frac{\ii\epsilon^{\mu\nu\rho}}{3!\sqrt{- h}}\epsilon_{abcd} E^b_\mu E^c_\nu E^d_\rho\,\sigma^a_{\alpha\dot\alpha}\, . 
	\ee 
	This condition reduces the number of independent components of $\kappa^\alpha, \bar\kappa^{\dot\alpha}$ from four to two real ones.
	
	The $\kappa$-invariance is indeed a fancy realization of a conventional local world-volume supersymmetry of the membrane which becomes manifest in the superembedding approach \cite{Bandos:1995zw,Sorokin:1999jx}. It can be used to put to zero half of the fermionic coordinates $\theta^\alpha(\xi), \bar\theta^{\dot\alpha}(\xi)$, 
	while the other half are dynamical world-volume fermionic fields playing the role of the Goldstinos for the partially spontaneously broken bulk supersymmetry. 
	In this sense, the membranes described by the action \eqref{SM} are $\frac12$-BPS objects.
	
	\item[Bulk super-diffeomorphism invariance.] Finally, \eqref{SM} is invariant under bulk super-diffeomorphisms.  When \eqref{SM} is included in the action of interacting system including also dynamical supergravity (like in our \eqref{fullS}), the superdiffeomorphism invariance is the gauge symmetry and  we may use it  to choose a membrane embedding such that the oscillations of the membrane in the transverse bosonic direction and along the fermionic direction look frozen, that is $y(\xi) = 0$ and $\theta(\xi)=0=\bar{\theta}(\xi)$. Therefore, in the interacting  system including dynamical (super)gravity we may assume the membrane to be static and located at $y_0=0$ without any loss of generality. In this gauge the dynamics of the membrane is encoded in the world-volume pull-backs of the bulk supergravity fields \cite{Bandos:2001jx,Bandos:2005ww}.
\end{description}

The super-membrane action \eqref{SM}-\eqref{susyNG} is also super-Weyl invariant. Fixing the super-Weyl invariance, performing the usual Weyl rescaling for passing to the Einstein frame  and putting to zero the fermionic fields, one reduces \eqref{SM} to the following bosonic action
\be\label{compSM}
\begin{aligned}
	S_{\rm M,\,bos}&=\,-2\int_\calc \d^3\xi\sqrt{- h}\,\; e^{\frac12\calk}\left| q_If^I(\phi)-p^I\calg_I (\phi)\right|+
	\\
	&\quad\,+q_I\int_\calc A_3^I-p^I\int_\calc\tilde A_{3I} \, .
\end{aligned}
\ee
From the Nambu--Goto term \eqref{susyNG} we immediately read the expression for the effective tension of the membrane
\be\label{tension}
T_{\text{M}}=2\, e^{\frac12 \calk} \left| q_If^I(\phi)-p^I\calg_I (\phi)\right|\, ,
\ee
which, rather than being a constant, depends on the values of the scalar fields on the membrane world-volume.

\begin{figure}[ht]
	\centering
	
	\includegraphics[width=7.5cm]{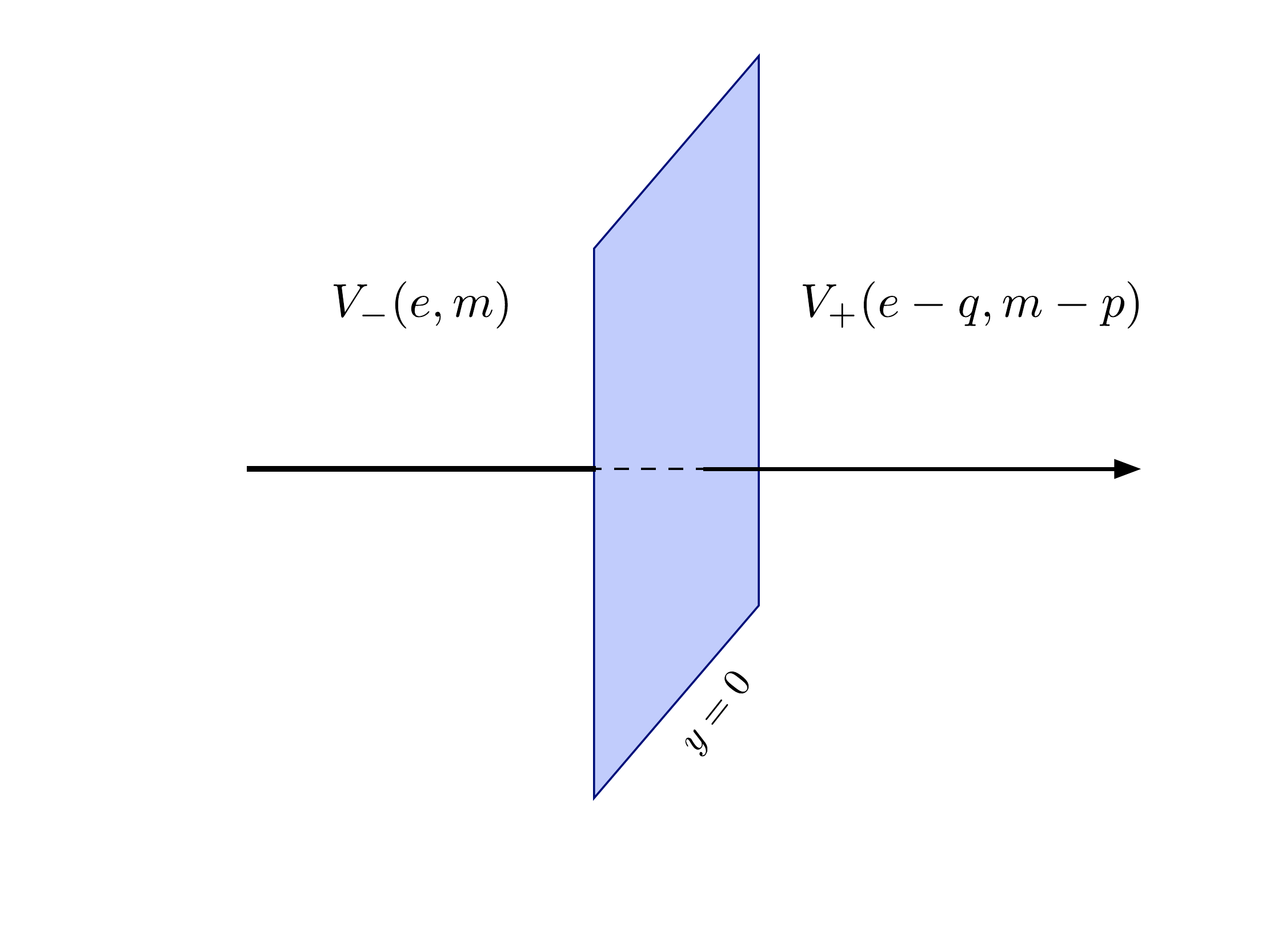} 
	\caption{\footnotesize The membrane divides the space-time into two regions. The different constants, $(e_{-I},m^I_-)$ on the left and $(e_{+I},m^I_+)$ on the right, determine different shapes of the potential in the two regions.}
	\label{fig:Membrane1}
\end{figure}

The full supersymmetric extension of the action \eqref{BTaction} is
\be
\label{fullS}
S= S_{\text{SG}} + S_{\text{M}}
\ee
whose bosonic components are \eqref{boscomp} and \eqref{compSM}. The coupling to membranes inevitably influences the equations of motion of the gauge three-forms which become
\be\label{eom3fM}
\begin{aligned}
	\d\Re\left(\calt^{IJ}{}^*\!\calf_{4J}\right)&=-\frac12p^I\delta_1(\calc)\,,
	\\
	\d\Re\left({\calg}_{IJ}\calt^{JK}{}^*\!\calf_{4K}\right)&=-\frac12q_I\delta_1(\calc)\,.
\end{aligned}
\ee
Here $\delta_1(\calc)$ is a delta-like one-form localized on the membrane world-volume $\calc$. Therefore, passing through the membrane, the quantized constants defined in \eqref{eom3f} jump as 
\be\label{fluxjump}
m^I_-\rightarrow m^I_+ \equiv m^I_--p^I\,,\quad e_{-I} \rightarrow e_{+I}\equiv e_{-I}-q_{I} \,, 
\ee
where $(e_{-I},m^I_-)$ are the parameters on the left of the membrane and $(e_{+I},m^I+)$ are those on the right. As a result, the membrane divides the space-time into two regions where the scalar fields feel different potentials, $V_-(e,m)$ on the left and $V_+(e-q,m-p)$ on the right (as depicted in Fig. \ref{fig:Membrane1}). In view of \eqref{potst} and \eqref{superp}, this means that we can define a superpotential over the whole space-time as
\be\label{Wmem}
\begin{split}
W(\phi,y)&=  e_{-I}f^I(\phi)-m_-^{I}\calg_I(\phi)\,-
\\
&\quad\,-\Theta(y)\left(q_If^I(\phi)-p^I\calg_I(\phi)\right) \, .
\end{split}
\ee

\begin{figure}[ht]
	\centering
	
	 \includegraphics[width=7.5cm]{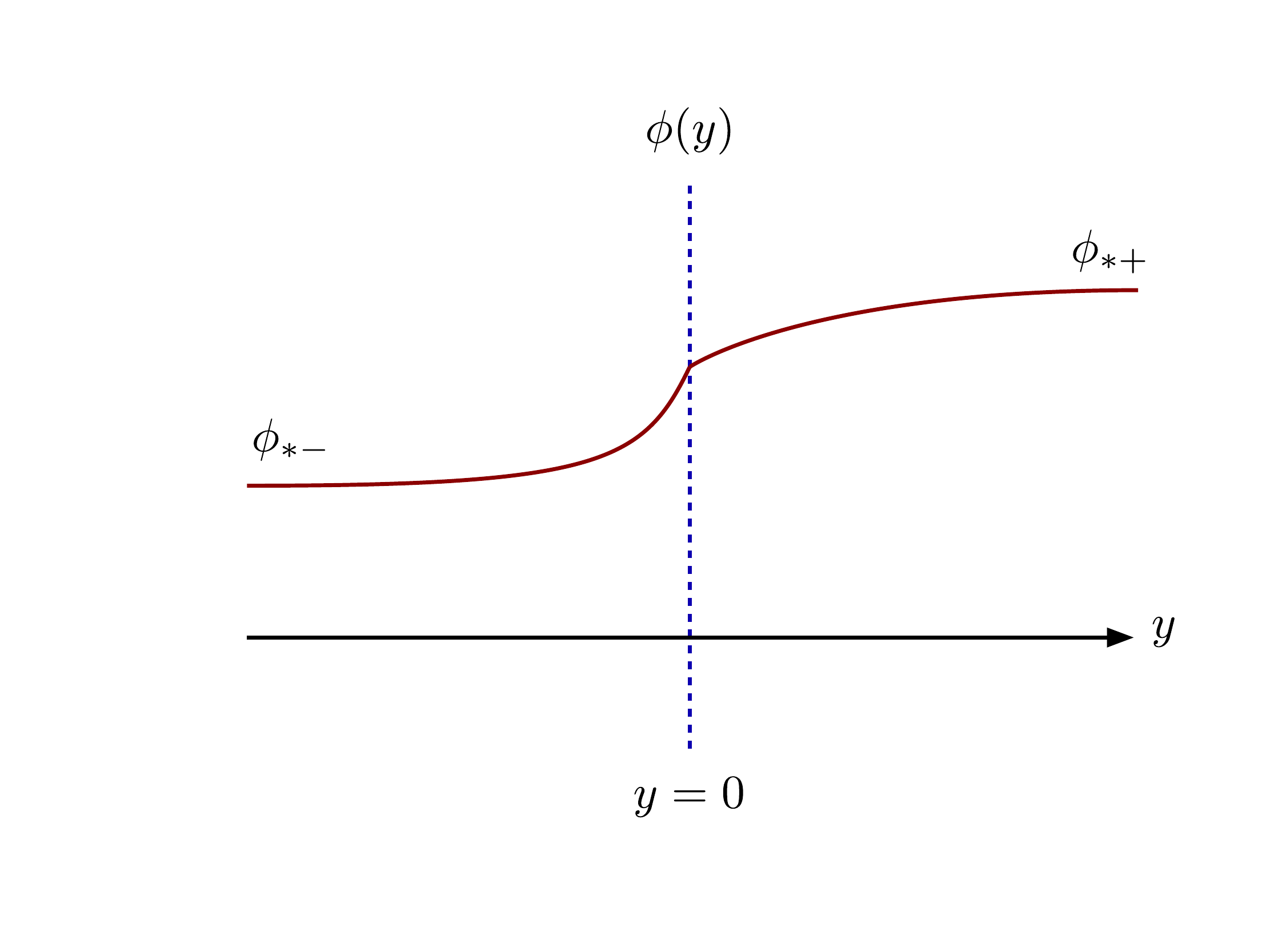}
	
	\caption{\footnotesize Here is depicted a domain wall-like solution interpolating between two supersymmetric vacua localized at $\phi_{*-}$ (on the left of the membrane) and $\phi_{*+}$ (on the right). Although $\phi(y)$ is continuous, its derivative might be discontinuous at $y=0$.}
	\label{fig:Membrane2}
\end{figure}


\section{Domain wall solutions and the flow equations}
\label{sec:DW}

We now pass to the study of domain wall solutions that connect the different vacua on the sides of the membrane. 
We consider a simple setting with a single flat membrane located at $y=0$ and we look for domain walls which are $\frac12$-BPS.

First, let us split the space-time coordinates $x^m$ in $x^\mu$, $\mu=0,1,2$ and $y\equiv x^3$, the latter being the coordinate transverse to the membrane. We consider the following domain wall ansatz for the space-time metric
\be\label{ds2=ansatz}
\d s^2=e^{2D(y)}\d x^\mu\d x_\mu+\d y^2 
\ee
and assume that the scalar fields $\phi^i$ depend only on the transverse coordinate $y$. 
The study of the domain wall solutions now proceeds along the same lines as in \cite{Cvetic:1992bf,Cvetic:1992st,Cvetic:1992sf,Cvetic:1993xe,Cvetic:1996vr,Ceresole:2006iq} for standard supergravity. By imposing that the supersymmetric variations of the fermions vanish for the metric ansatz \eqref{ds2=ansatz}, we arrive at the following equations
\begin{subequations}\label{floweq1}
	\begin{align}
	\dot \phi^i &= e^{\frac 12\calk(\phi,\bar\phi)+i\vartheta(y)}K^{i\bar\jmath}(\overline W_{\bar\jmath}+K_{\bar\jmath} \overline W),\label{susy22}\\
	\dot D &= - e^{\frac 12\calk(\phi,\bar\phi)}|W|,\label{susy11}
	\\
	\dot\vartheta &=-\Im\left(\dot\phi^i K_i\right), 	\label{thetaflow}
	\end{align}
\end{subequations}
where the dot corresponds to the derivative with respect to $y$ and $\vartheta(y)= \vartheta(\phi(y),\bar{\phi}(y))$ is the phase of $W$, that is $W=e^{i\vartheta}|W|$. Here $W$, although it generically depends on the field strengths $\calf_{4I}$, reduces to \eqref{Wmem} when setting the gauge three-forms on-shell.

The equations \eqref{floweq1} are \emph{flow equations} that describe how the scalars, the warp factor $D(y)$ and the phase of the superpotential vary along the direction transverse to the membrane and determine the domain wall solution. In order to better characterize the flow, we can introduce a \emph{`flowing' covariantly holomorphic superpotential} \cite{Ceresole:2006iq} 
\begin{eqnarray}\label{jumpc}
\calz(\phi,y)& \equiv & e^{\frac12 \calk(\phi,\bar\phi)}W\nonumber\\
&= & e^{\frac12 \calk(\phi,\bar\phi)}\left[\Theta(y)W_+(\phi)+\Theta(-y)W_-(\phi)\right] \, . 
\end{eqnarray}
With the use of \eqref{jumpc} the equations \eqref{floweq1} are recast to the form 
\begin{subequations}\label{floweq}
	\begin{align}
	\dot \phi^i &= 2K^{i\bar\jmath}\,\del_{\bar\jmath}|\calz|,\label{susy2}\\
	\dot D &= - |\calz|,\label{susy1}\\
	\dot\vartheta &=-\Im\left(\dot\phi^i K_i\right).	\label{thetaflowb}
	\end{align}
\end{subequations}

From \eqref{susy2}, we see that the fixed points for the flow of the scalars are those for which $\del_{\bar \jmath} |\calz|=0$ (or, equivalently, $D_{\bar \jmath} \bar W =0$). Then, the domain wall solution interpolates between two supersymmetric vacua, that are specified by the field configurations $\phi_{*-}^i$ on the left, reached in the limit $y \to -\infty$, and $\phi_{*+}^i$ on the right to the membrane, reached as $y \to +\infty$ (see Fig. \ref{fig:Membrane2} for an example of the flow of the scalars). These are generically AdS vacua since asymptotically from \eqref{susy1} it follows that $D_{\pm} = - |\calz_{*\pm}| y$ (where $\calz_{*\pm} \equiv \lim_{y\to \pm \infty} \calz(y)$), which correspond to AdS spaces with radii $1/\calz_{*\pm} $.

In order to identify a $c$-function, let us notice that, combining the membrane equations of motion with the flow equations \eqref{floweq}, we get
\be
\label{diffcalz2}
\begin{split}
\frac{\d |\calz|}{\d y}&= K_{i\bar\jmath}\dot\phi^i \dot{\bar\phi}^{\bar\jmath} +\frac12 T_{\rm M}\,\delta(y)
\\
&=4K^{i\bar\jmath}\del_i|\calz|\del_{\bar\jmath}|\calz|+\frac12T_{\rm M}\delta(y)\geq 0 \,.
\end{split}
\ee
Hence, $|\calz(y)|=-\dot D(y)$ is a monotonic increasing function and the flow is towards the direction where $|\calz|$ increases. In the following, we shall assume $|\calz|_{y=+\infty} > |\calz|_{y=-\infty}$, the other choice being obtained just by flipping $y \to -y$. Then, the flow starts from a supersymmetric vacuum $\phi_{*-}$ to the left, at $y=-\infty$, crosses the membrane and a new supersymmetric vacuum $\phi_{*+}$ is reached asymptotically to the right.

From \eqref{jumpc} one recovers the tension of the membrane. In fact, close to the membrane \be\label{calzjump}
\begin{split}
 \Delta\calz &\equiv\lim_{\varepsilon\rightarrow  0}\left(\calz|_{y=\varepsilon}-\calz|_{y=-\varepsilon}\right)=
 \\
 &=-e^{\frac12 \calk}\left(q_If^I-p^I\calg_I\right)|_{y=0}  
\end{split}
\ee
and \eqref{tension} can be expressed as
\be\label{effTM}
T_{\rm M}\equiv 2\,e^{\frac12 \calk}\left|q_If^I-p^I\calg_I\right|_{y=0}=2|\Delta\calz|\,.
\ee
However, we may also compute the tension of the domain wall which, although localized mainly close to the membrane, extends over the whole space-time. To this aim, we first plug the metric ansatz \eqref{ds2=ansatz} into the full action \eqref{fullS} and then rewrite \eqref{fullS} in the \emph{BPS-form}
\be\label{redS}
\begin{aligned}
	S_{\rm red}=&\,\int\d^3 x\int\d y\, e^{3D}\bigg[3\big(\dot D +|\calz|\big)^2-
	\\
	&-K_{i\bar\jmath}\big(\dot\phi^i-2K^{i\bar k}\del_{\bar k}|\calz|\big)\big(\dot{\bar\phi}^{\bar\jmath}-2K^{l\bar \jmath}\del_l|\calz|\big)\bigg]-
	\\
	&-2\int\d^3 x\left[\big( e^{3D}|\calz|)|_{y=+\infty}-\big(e^{3D}|\calz|\big)|_{y=-\infty}\right].
\end{aligned}
\ee
Its on-shell value is precisely the energy of the solitonic solution, that is the tension of the domain wall. Noticing that the first two lines contain 
\eqref{susy22}, \eqref{susy11} and vanish on-shell, we get 
\be\label{DWtension}
T_{\rm DW}=2\big(|\calz|_{y=+\infty}-|\calz|_{y=-\infty}\big)\,.
\ee
It is important to stress that, generically, the tension of the membrane and of the domain wall are different, being related via
\be\label{DWtension2}
\begin{split}
T_{\rm DW}&=2\big(|\calz|_{y=+\infty}-\lim_{\varepsilon\rightarrow 0}|\calz|_{y=\varepsilon}\big)\,+
\\
&\quad\,+2\big(\lim_{\varepsilon\rightarrow 0}|\calz|_{y=-\varepsilon}-|\calz|_{y=-\infty}\big)+ T_{\rm M}\,.
\end{split}
\ee
They coincide only if the $|\calz|$ is just a constant on the two sides. This is the case, for example, of the \emph{thin-wall}  approximation, but for \emph{thick} walls $T_{\rm DM} > T_{\rm M}$ holds strictly.

\section{Conclusion and outlook}

In this contribution we have reviewed  the main results of \cite{Farakos:2017jme} and \cite{Bandos:2018gjp} on the construction of 4D supergravities containing gauge three-forms and membranes. 
These theories provide an effective supergravity description of type II string compactifications 
on Calabi--Yau manifolds with Ramond-Ramond fluxes. 
In particular, in \cite{Farakos:2017jme,Bandos:2018gjp} it has been shown that the scalar potential obtained from compactifications of Type IIA string theory \cite{Grimm:2004ua,Louis:2002ny,Bielleman:2015ina,Herraez:2018vae,Carta:2016ynn} is  retrieved from \eqref{potst} and that the tension of the membrane \eqref{tension} is equal to that of membranes obtained from wrapping higher dimensional branes over special Lagrangian cycles in the Calabi--Yau. 

Setting the gauge three-forms on-shell results in distinct potentials in each of the space-time regions separated by the membranes and standard supergravity theories are recovered in each region. In turn, this implies that different regions have different vacua characterized by different values of the cosmological constant. All these vacua might be connected by domain wall solutions, and we have studied the equations that regulate the flow of the scalar fields and the metric warp factor in such solutions.

In a more general context, the results reviewed above furnish an effective field-theoretical ground 
for studying a variety of problems in string-inspired phenomenological and cosmological model building. 
However, even though our approach is quite general, the application to more involved string compactifications requires further work. 
First of all, in the type IIA models reviewed in this article the tadpole condition does not directly concern the internal fluxes subject to the dualization.   
In more general IIA compactifications, for instance with a non-trivial $H_3$-flux, the tadpole condition would become relevant for the dualization procedure. The same is true for type IIB orientifold compactifications, which have flux induced superpotential \cite{Gukov:1999ya,Gukov:1999gr,Taylor:1999ii,Giddings:2001yu,Grimm:2004uq}  compatible with our general framework too. Also in these cases a non-trivial tadpole condition should be appropriately taken into account \cite{Bandos:2018gjp}. 

Another aspect that deserves further study is the inclusion of the open-string sector in the effective theory, which may be naturally incorporated in a three-form formulation  \cite{Grimm:2011dx,Kerstan:2011dy,Carta:2016ynn,Herraez:2018vae}. It would be interesting to revisit this point in the manifestly supersymmetric framework reviewed here. Related questions concern its applications to M-theory and  F-theory compactifications, which can be considered as strong coupling limits of type IIA and IIB compactifications with backreacting branes, see for instance \cite{Acharya:2004qe,Denef:2008wq} for reviews. 

Finally, in the presence of spontaneously broken supersymmetry, our formulation should be related, at low energies, to models with non-linearly realized local supersymmetry as the ones discussed in  \cite{Farakos:2016hly,Buchbinder:2017vnb}. It would be interesting to elucidate this relation. This general framework can also be used to construct and study supersymmetric extensions of other models based on gauge three-forms, as for instance those discussed in \cite{Dvali:2005an,Dvali:2004tma,Dvali:2005zk,Dvali:2013cpa,Dvali:2016uhn,Dvali:2016eay}.

\bibliography{allbibtex}

\bibliographystyle{prop2015}

\end{document}